\documentclass[sigconf]{acmart}

\usepackage{url}
\usepackage{code}
\usepackage{graphicx}
\usepackage{balance}
\usepackage{multirow}
\usepackage{multicol}
\usepackage{listings}
\usepackage{booktabs}
\usepackage{subcaption}
\usepackage{cleveref}
\usepackage{xspace}
\usepackage[most]{tcolorbox}
\usepackage{pifont}
\usepackage{fontawesome} 
\hypersetup{draft}
\usepackage{hyperref}
\usepackage{array}
\usepackage{adjustbox}
\usepackage{enumitem}
\usepackage{tabularx}

\definecolor{mycolor}{RGB}{119,195,236}

\newtcolorbox{highlight}[1][]{
  colback=mycolor!20!white,
  colframe=mycolor!80!black,
  fonttitle=\bfseries,
  coltitle=mycolor!80!black,
  colbacktitle=mycolor!20!white,
  boxrule=1pt,
  arc=3pt,
  outer arc=3pt,
  boxsep=0pt,
  left=10pt,
  right=10pt,
  top=6pt,
  bottom=6pt,
  toptitle=2pt,
  bottomtitle=2pt,
  lefttitle=2pt,
  righttitle=2pt,
  titlerule=0pt,
  attach boxed title to top left={yshift=-\tcboxedtitleheight/2, xshift=2mm},
  boxed title style={arc=3pt, outer arc=3pt},
}

\lstdefinestyle{code}{
  language=C,
  frame=single,
  backgroundcolor=\color{white},
  basicstyle=\footnotesize\ttfamily,
  keywordstyle=\color{blue}\bfseries,
  commentstyle=\color{teal},
  stringstyle=\color{red},
  numbers=left,
  numberstyle=\tiny\color{gray},
  stepnumber=1,
  numbersep=5pt,
  showspaces=false,
  showstringspaces=false,
  showtabs=false,
  tabsize=2,
  captionpos=b,
  breaklines=true,
  breakatwhitespace=true,
  escapeinside={\%*}{*)},
  morekeywords={*,...}
}

\newcommand{\instruction}[0]{\faInfo}
\newcommand{\codeFile}[0]{\faFolderOpenO}
\newcommand{\codeDiff}[0]{\faCodeFork}
\newcommand{\codeDesc}[0]{\faCopy}
\newcommand{\codeDescDiff}[0]{\faColumns}

\newcommand{\tests}[0]{\faFileCodeO}
\newcommand{\fullSpec}[0]{\faFileTextO}
\newcommand{\relevantSpec}[0]{\faListAlt} 
\newcommand{\constraints}[0]{\faCheckSquareO}
\newcommand{\bugClasses}[0]{\faBug}
\newcommand{\testDesc}[0]{\faAlignLeft}
\newcommand{\testCode}[0]{\faCode}
\newcommand{\guidelines}[0]{\faPencilSquareO}

\graphicspath{{Figures/}}

\begin{document}

\title{\tool: Differential testing with LLMs using Natural Language Specifications and Code Artifacts}


\newcommand{\mr}[2]{\multirow{#1}{*}{#2}}
\newcommand{\mc}[3]{\multicolumn{#1}{#2}{#3}}

\newcommand{\todo}[1]{\textbf{\textcolor{red}{TODO: #1 }}}
\newcommand{\update}[1]{{\textcolor{blue}{#1 }}}
\newcommand{\nr}[1]{\textbf{\textcolor{teal}{NR: #1 }}}
\newcommand{\clg}[1]{\textbf{\textcolor{purple}{CLG: #1 }}}
\newcommand{\sarah}[1]{\textbf{\textcolor{blue}{Sarah: #1 }}}
\newcommand{\evg}[1]{\textbf{\textcolor{orange}{EVG: #1 }}}
\newcommand{\hg}[1]{\textbf{\textcolor{brown}{HG: #1 }}}

\newcommand{\tool}{\textsc{DiffSpec}\xspace}

\author{Nikitha Rao}
\email{nikitharao@cmu.edu}
\affiliation{%
  \institution{Carnegie Mellon University}
  \city{Pittsburgh}\state{PA}  \country{USA}
}

\author{Elizabeth Gilbert}
\email{evgilber@andrew.cmu.edu}
\affiliation{%
  \institution{Carnegie Mellon University}
  \city{Pittsburgh}\state{PA}  \country{USA}
}

\author{Harrison Green}
\email{harrisog@andrew.cmu.edu}
\affiliation{%
  \institution{Carnegie Mellon University}
  \city{Pittsburgh}\state{PA}  \country{USA}
}

\author{Tahina Ramananandro}
\email{taramana@microsoft.com}
\affiliation{%
  \institution{Microsoft Research}
  \city{Redmond}\state{WA}  \country{USA}
}

\author{Nikhil Swamy}
\email{nswamy@microsoft.com}
\affiliation{%
  \institution{Microsoft Research}
  \city{Redmond}\state{WA}  \country{USA}
}

\author{Claire Le Goues}
\email{legoues@cmu.edu}
\affiliation{%
  \institution{Carnegie Mellon University}
  \city{Pittsburgh}\state{PA}  \country{USA}
}

\author{Sarah Fakhoury}
\email{sfakhoury@microsoft.com}
\affiliation{%
  \institution{Microsoft Research}
  \city{Redmond}\state{WA}  \country{USA}
}

\renewcommand{\shortauthors}{Rao et al.}

\begin{abstract}

Differential testing can be an effective way to find bugs in software systems with multiple implementations that conform to the same specification, like compilers, network protocol parsers, or language runtimes. 
Specifications for such systems are often standardized in natural language documents, like Instruction Set Architecture (ISA) specifications or IETF RFC's. 
Large Language Models (LLMs) have demonstrated potential in both generating tests and handling large volumes of natural language text, making them well-suited for analyzing artifacts like specification documents, bug reports, and code implementations. 
In this work, we leverage natural language and code artifacts to guide LLMs to generate targeted tests that highlight meaningful behavioral differences between implementations, including those corresponding to bugs. We introduce \tool, a framework for generating differential tests with LLMs using prompt chaining. We demonstrate \tool's efficacy on two different (extensively tested) systems, eBPF runtimes and Wasm validators. Using \tool, we generated 1901 differentiating tests, uncovering at least four distinct and confirmed bugs in eBPF, including a kernel memory leak, inconsistent behavior in jump instructions, undefined behavior when using the stack pointer, and tests with infinite loops that hang the verifier in ebpf-for-windows.  We also found 299 differentiating tests in Wasm validators pointing to two confirmed and fixed bugs.

\end{abstract}



\setcopyright{none} 
\settopmatter{printacmref=false} 

\renewcommand\footnotetextcopyrightpermission[1]{}

\maketitle

\section{Introduction}

Differential testing is an approach for automatically generating potentially bug-finding tests for applications that correspond to multiple implementations of the same functionality~\cite{mckeeman1998differential}.  
The key idea is to test two or more different systems (or two different versions of the same system) that should behave the same way under the same conditions on the same inputs.  
If their output behavior differs, it is likely that at least one of the implementations is incorrect. 

Differential testing has shown significant success especially in testing language implementations, such as uncovering bugs in C compilers~\cite{yang2011finding, le2014compiler} or browser engines, for example revealing inconsistencies in JavaScript interpreters and JIT compilers~\cite{bernhard2022jit}. It also can be useful for testing cross-platform consistency (i.e., the same system across different configurations or operating systems)~\cite{DiffDroid} or versions (as in regression testing)~\cite{diff-rest}. 

Generating tests that specifically target differences between two versions of a program is especially challenging, as it 
involves simultaneously searching the vast input space of two programs to find rare inputs that trigger often subtle  discrepancies~\cite{mckeeman1998differential}. Existing approaches to find such tests limit the possible search space by borrowing techniques from symbolic execution~\cite{rutledge2022automating}, guided semantic aware program generation~\cite{kapus2017automatic}, type aware mutations~\cite{jay2018structured}, and code coverage optimizations~\cite{chen2016coverage}. While some approaches leverage semantic and syntactic properties of the code or use information from static analysis tools, they are significantly limited in their ability to harness the wealth of information available from natural language artifacts.

Large Language Models (LLMs) excel at extracting and understanding information from large amounts of natural language text, enabling a wide variety of tasks such as program comprehension, bug localization, and software testing. Recent research has shown significant promise in leveraging LLMs to enhance various testing techniques. For instance, LLMs have been used to generate more effective mutations in mutation testing~\cite{deng2023large}, to create higher-quality unit tests~\cite{rao2023cat, alshahwan2024automated, li2023nuances, liu2024llm}, and to improve fuzzing methods by producing diverse and targeted inputs~\cite{xia2024fuzz4all}.

In this work, we propose a differential testing technique that leverages natural language and code artifacts describing the software systems under test to inform and prompt an LLM to produce effective, and targeted, differential tests. The extracted information describes the system specification, its source code implementation, and historical bug information.

We realize this intuition in \tool, a general approach to differential testing of multiple systems implemented with respect to a documented specification, and with functionality that can be decomposed into testable units. 
\tool is well-suited for testing systems that correspond to, or integrally include, language compilers, runtimes, and verification systems, like network protocol parsers  or JVM or EVM or web browsers. 
This is the predominant domain for differential testing applications in research~\cite{compiler-driven, ssl-diff, browser-diff} and practice~\cite{cedar}. 
Such systems are typically associated with comprehensive language specification documentation, like the Instruction Set Architecture (ISA) specification associated with eBPF~\cite{bpf-spec}, or the WebAssembly (Wasm) language specification~\cite{wasm-spec}.
Moreover, testing these types of systems can be naturally decomposed into testing language instructions or subsets thereof.  

We demonstrate \tool on various implementations of Wasm and eBPF runtimes, which both have rich evolving natural language artifacts that \tool can leverage, and are widely used in practice. Both runtimes vary in domain, the type of contextual information that our approach must extract from the natural language artifacts, and the format and language of the tests to be generated (see Section~\ref{sec:systems}), demonstrating the generalizability of \tool. 

Using \tool, we found 299 differentiating tests across four different implementations of Wasm validators.  Upon manual analysis, we found that these point to at least two bugs which includes a type mismatch and cast of out-of-bounds. These bugs were reported to the maintainers of Wasm and have now been fixed. We also generated 1901 differentiating tests, that helped discover at least four distinct bugs across three different implementations of eBPF runtimes. These include a kernel memory leak, inconsistent behavior in jump instructions, undefined behavior when using the stack pointer, and tests with infinite loops that hang the verifier in ebpf-for-windows. These bugs were confirmed by the contributors of eBPF and issues have been filed for them. 

\noindent In summary, we make the following contributions:
\begin{itemize}[leftmargin=*,nolistsep]
    \item We release \tool, the first differential test generation framework that uses natural language specifications and code artifacts to generate tests.
    \item We demonstrate the generalizability of the framework by evaluating it on two real-world systems, eBPF and Wasm. 
    \item We expand the existing test suites for both eBPF and Wasm, and contribute to open source. 
    \item We find bugs in the implementations of both eBPF and Wasm runtimes, which have been reported to the maintainers of the projects. The Wasm bugs have been fixed since we reported them.
    
\end{itemize}

\section{Illustrative Example}
\label{sec:example}

This section presents an example illustrating how \tool uses natural language specifications and code artifacts to generate tests for eBPF.  First, given a natural language specification document~\cite{}, \tool extracts a list of instructions in the underlying language, along with the corresponding constraints for each instruction. For example, constraints extracted for the \texttt{RSH} instruction include:
\begin{enumerate}[leftmargin=*,nolistsep]
    \item The RSH instruction performs a right shift operation. The destination register (dst) is shifted right by the number of bits specified in the source operand (src or imm). 
    \item The source operand can come from either the src register (if the source bit in the opcode is set to X) or the immediate value (if the source bit in the opcode is set to K).
    \item For ALU: \{RSH, K, ALU\} means $dst = (u32)(dst >> imm)$ and \{RSH, X, ALU\} means $dst = (u32)(dst >> src)$.
\end{enumerate}

Next, for each considered instruction \tool extracts the implementation of that instruction from each of the two codebases corresponding to the systems under test.   
For example, \tool extracts an implementation of \texttt{RSH} from the source code for eBPF in linux ARM 32, a subset of which includes:

\begin{lstlisting}[caption=RSH source code extracted from the bpf implementation from linux arm 32 implementation, label=lst:rsh-code]
/* ... */
/* dst = dst >> src */
case BPF_ALU | BPF_RSH | BPF_X:
  case BPF_ALU64 | BPF_RSH | BPF_X:
    switch (BPF_SRC(code)) {
      case BPF_X: /* Shift right by variable */
        emit_a32_alu_r64(is64,dst,src,ctx,BPF_OP(code));
        break;
      case BPF_K:	/* Shift right by immediate value */
			if (unlikely(imm > 31))
				return -EINVAL;
/* ...continues, elided... */
\end{lstlisting}

Given two code snippets, \tool then reasons about the implementation differences, such as ``Checking of Immediate Value: The first implementation checks if the immediate value is greater than 31 or 63 for 32-bit and 64-bit operations respectively. The second implementation does not perform this check."

\tool additionally looks at historical bugs to generate a set of bug classes  to guide test generation. For example, ``Shift Operation Bug: These bugs occur when the JIT compiler incorrectly handles shift operations, especially when the shift amount is zero. Incorrect shift operations can lead to unexpected results, or in worst cases, hang the kernel.''

Using all extracted context, \tool first generates a set of test descriptions that detail what the test should check for. This along with a set of hand written guidelines that provides instructions on what makes a valid test, is then used to generate the test code. Here is an example of the test description along with the corresponding test code generated by \tool.
\begin{lstlisting}[caption=Generated test code, label=lst:rsh-test-code]
// Test with a zero-shift count. 
// Check for edge cases where the shift count is zero. 
// The source value should remain unchanged.
-- asm
mov %r0, 0x12345678
rsh %r0, 0
exit
-- result
0x12345678
\end{lstlisting}

Interestingly, there was a historical bug in the linux implementation of BPF.\footnote{
https://git.kernel.org/pub/scm/linux/kernel/git/torvalds/linux.git/commit/?id=bb9562cf5c67
} The issue titled ``arm, bpf: Fix bugs with ALU64 {RSH, ARSH} BPF\_K shift by 0'' describes how ``The current arm BPF JIT does not correctly compile RSH or ARSH when the immediate shift amount is 0...'' The test generated by \tool would have caught this issue, preemptively.

\section{Approach}
\label{sec:overview}

 \begin{figure*}
    \center
     \includegraphics[width=\textwidth,trim={1.2cm 5cm 3cm 3.5cm}, clip]{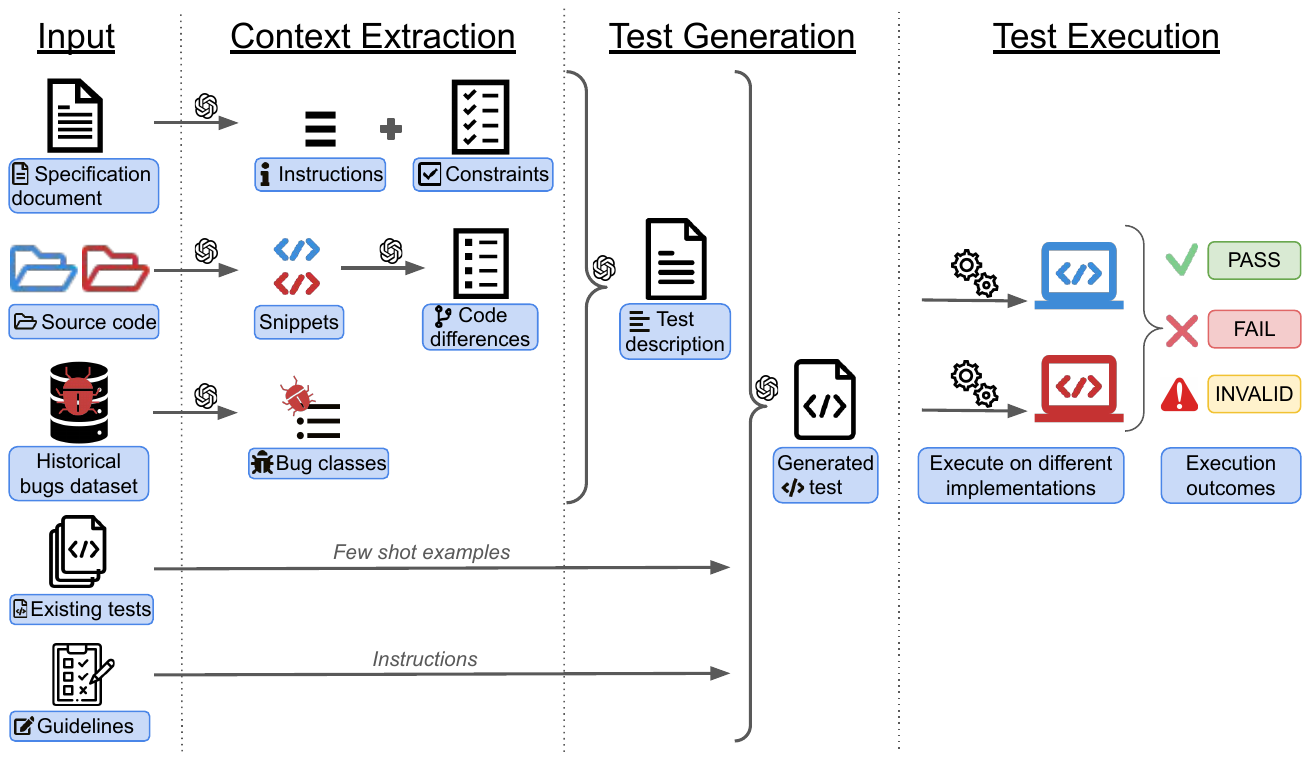}
    \caption{Approach overview. \tool extracts relevant context from natural language and code artifacts by prompting an LLM, covering: instructions and constraints, source code and tests, and historical bug information. \tool then follows a two-step process: (1) it generates test descriptions using the extracted context, and then (2) uses the test description along with few-shot examples of human written tests along with a set of human written guidelines to generate actual test code. Generated tests are executed on different implementations of the specification, seeking those that are potentially differentiating.}
 
    \label{fig:overview}
  \end{figure*}

Figure~\ref{fig:overview} provides an overview of how \tool uses LLMs to generate tests using natural language specifications and code artifacts. The core components of \tool are:
\begin{itemize}[leftmargin=*,nolistsep]
\item \textbf{Extracting relevant context} from software artifacts (Section~\ref{sec:extract-context}) by prompting an LLM. The information includes but is not limited to, for each instruction: relevant constraints; source code snippets from the tested implementations; and bug information.  
\item \textbf{Generating tests} (Section~\ref{sec:test-generation}) using the extracted information as context to  prompt an LLM to generate natural language descriptions of test cases.
This description, along with relevant human-written tests (used as few-shot examples) and a set of curated instructions guide the model to generate actual test inputs.
\item \textbf{Evaluating the systems} under test. Finally, we execute generated tests on the implementations under test with an evaluation harness. 
A generated test is \emph{differentiating} if different implementations produce different outputs for the same test input.  Not all such tests correspond to defects, but they serve as a key starting point to identify potential problems in either the implementation(s), or the specification.
\end{itemize}

The rest of this section describes the first two components in detail. Evaluating the systems under test is straightforward conceptually; we provide implementation specifics with respect to our evaluation systems in Section~\ref{sec:systems}. Full prompts are available in the artifact.\footnote{\url{https://doi.org/10.5281/zenodo.13756137}} 

\subsection{Extraction of relevant context from artifacts} 
\label{sec:extract-context}

Software systems are associated with many artifacts that encode desired and expected behavior as well as potential incorrect behavior.  \tool leverages these artifacts to guide the LLM to generate targeted tests that highlight informative behavioral differences between different implementations. 
The considered sources of information include (1) a system \emph{specification}, which are typically semi-structured natural language documents describing what the multiple systems under test \emph{should} do, (2) system \emph{implementations} of the functionality under test, (3) system \emph{tests} provided by developers, and (4) \emph{historical bug information}, drawn from a bug report database or expert knowledge, that can provide clues about common failure modes. This results in the following types of information:

\vspace{1ex}
\noindent\emph{Instructions (\instruction) and Constraints (\constraints)}
Given a natural language specification document for a given system specification, \tool first prompts the LLM to extract all instructions in the implemented language (\instruction).
Specification documents describe the specified behavior of each instruction in a language, including constraints (\constraints) on correct implementation. \tool therefore prompts an LLM to extract this information for each instruction, again from the specification document.  
Language specifications can be dozens to hundreds of pages long; extracting only the text that describes each instruction allows subsequent prompts to be specific, and overcomes LLM context window limitations.

\vspace{1ex}
\noindent\emph{Implementation code (\codeFile)}
Code-based artifacts can provide useful guidance to differential test generation. First,
 for each instruction, \tool uses its extracted constraints as a guide to identify and extract relevant source code snippets from each implementation.
Although implementations of a given specification vary, they are expected by construction to be semantically equivalent. 
We hypothesize that implementation differences can be used to guide the LLM to generate tests that may result in differential behavior. In order to get these differences we first extract the relevant code pertaining to the given instruction by prompting the LLM with the source code files and the list of constraints. \tool then uses the extracted code snippets from each implementation to prompt the LLM to reason about differences, producing a list in natural language (\codeDiff). 

LLMs have been shown effective at generating summaries, including code summaries~\cite{sun2024source}. Inspired by the self-debug~\cite{selfdebug} work, which uses an LLM to explain the code and compares the natural language summary to the problem description to find bugs, we explore an alternative approach to extract information from the code snippets by having the LLM describe what the code is doing in natural language. We then use the code descriptions ((\codeDesc) of both the implementations to guide the test generation process. We take this one step further and also have the LLM reason about the differences in the two descriptions (\codeDescDiff). We compare the efficacy of these approaches in Table~\ref{tab:results-ebpf}.

\vspace{1ex}
\noindent\emph{Test code (\tests)} Mature software systems include human-written test suites that exercise functionality that \tool also targets, like specific instruction implementations.
\tool generates a mapping between each instruction and any human written tests for it by providing the LLM lists of instructions, and test file names. The mapping provides two main benefits. First, it helps identify instructions that don't have a corresponding test, highlighting gaps. Second, the mapping provides examples for test generation.

\vspace{1ex}
\noindent\emph{Bug categories (\bugClasses)}
Finally, historical bug information can help provide useful context on common edge cases or failure modes. \tool can make use of multiple sources of such information including bug reports, prior empirical studies, and expert knowledge; it queries an LLM to distill the information accordingly. 

\subsection{Test Generation Framework}
\label{sec:test-generation}

\tool uses the context extracted from the natural language and code artifacts to generate differential tests.   It does this in two phases:

\vspace{1ex}
\noindent\emph{Generating test descriptions (\testDesc)}
\tool incorporates the specifications and constraints for each tested instruction, along with all the additional extracted context, to prompt the LLM to generate a configurable number (we use 10 in our experiments) of natural language descriptions for what a test should do. This step is repeated for every combination of extracted code difference and bug category.

\vspace{1ex}
\noindent\emph{Generating tests (\testCode)} Given a natural language test description, \tool next prompts the LLM to convert these into executable tests. The generated tests include expected output.  We also include an optional set of guidelines (\guidelines) for valid tests for the system, provided by a human expert; we evaluate its contribution to test validity in Section~\ref{sec:ablation}.  
In our analysis, the two-phase approach is generally more effective for generating syntactically valid code than a single-phase approach (See Table~\ref{tab:results-ebpf}).

\section{Systems Under test}
\label{sec:systems}

\begin{table*}[t!]
\small
  \begin{center}
    \caption{Selection of bug categories and descriptions for eBPF and Wasm}
    
    \label{tab:bug_categories}
    
    \begin{tabularx}{\textwidth}{p{1cm}X}

    \toprule
    \textbf{Target} & \textbf{\textit{Category} - Description} \\
    \midrule

        \textbf{eBPF} & \textbf{\textit{Instruction Encoding}} - Incorrect assembly code generation by the eBPF JIT compiler, involving incorrect opcodes, registers, or misinterpreting eBPF instructions.\\
       & \textbf{\textit{Stack Layout}} - Incorrect stack frame setup or teardown, causing memory corruption.\\
       & \textbf{\textit{Shift Operation}} - Incorrect handling of shift operations, e.g. shift by 0 errors, leading to unexpected results, or hanging the kernel.\\
       & \textbf{\textit{Register Handling}} - Incorrect usage, saving, or restoration of CPU registers. \\
      
       & \textbf{\textit{Endianness Conversion}} - Incorrect conversions from big and little endian representations.\\
    
        \midrule

        \textbf{Wasm}& \textbf{\textit{Branch Target Resolution}} - Branching instructions rely on specifying labels to determine the branch target. A bug in resolving these labels can cause the control flow to jump to the wrong location, leading to unexpected behavior. \\
       
       & \textbf{\textit{Block Nesting}} - Failure to correctly manage block nested structures, leading to incorrect flow of control. For instance, a \texttt{br\_table} instruction that incorrectly interprets the depth of nested blocks can cause the control flow to exit the wrong block or loop.\\    
           
        & \textbf{\textit{Type Mismatch in Control Flow}} -  Control flow instructions must adhere to specific type constraints. otherwise leading to runtime type errors. \\    
         
       & \textbf{\textit{Control Flow Across Module Boundaries}} - Wasm modules can import and export functions, and bugs can occur if control flow instructions don't correctly handle calls or returns across module boundaries. \\

   \bottomrule      
    \end{tabularx}

  \end{center}
\end{table*}

\tool is a framework that can apply to many different systems for which differential testing is appropriate.  For the purposes of evaluation, we apply it to two complex real-world problem specifications and several associated implementations: the extended Berkeley Packet Filter (eBPF) (Section~\ref{sec:ebpf}),
a kernel-extension framework that safely runs custom bytecode programs; and WebAssembly (Wasm) (Section~\ref{sec:wasm}), a portable bytecode language and compilation target originally designed for browser-based applications but with broad application beyond. 
This Section describes these systems, with particular focus on details relevant to our evaluation.

\subsection{extended Berkeley Packet Filter (eBPF)}
\label{sec:ebpf}

\noindent\emph{Background.} The extended Berkeley Packet Filter,  or eBPF, is a kernel-extension framework originally integrated into the Linux kernel at version 3.18~\cite{linux-kernel-ebpf}.
It allows developers to safely run custom bytecode programs inside the kernel, without inserting risky modules or modifying the kernel itself.  
A key component of eBPF ecosystem is the verifier, which ensures safety properties (like memory safety, crash-freedom, or termination) of user-defined extensions.   
Support for eBPF framework has been implemented for multiple runtime environments and architectures (including but not limited to x86\_64, ARM, Risc-V). 
There furthermore exists several user-space eBPF runtime implementations~\cite{ebpf} that extend its reach beyond kernel-level interactions. 

\noindent\emph{Tested Implementations and Code artifacts.} We test three eBPF runtimes in our experiments:
(1) Linux Kernel via Libbpf~\cite{libbpf}, a user-space library that simplifies the use of eBPF programs in the Linux kernel. 
(2) Userspace BPF (uBPF)~\cite{ubpf}, a lightweight, user-space implementation of eBPF. 
(2) eBPF for Windows via bpf2c~\cite{ebpf-for-windows}, which Microsoft has introduced into the Windows ecosystem, again providing a user-space eBPF runtime.

We take the 206 human-written tests for \texttt{bpf\_conformance} for the test artifacts. 

\vspace{1ex}
\noindent\emph{Natural language Artifacts.}
All runtime implementations of eBPF must conform to the eBPF Instruction Set Architecture~\cite{bpf-spec}, standardized and documented through the IETF, and the authoritative source for the specification standard. We use the BPF ISA as the specification document for testing; \tool extracts a total of 34 instructions from the ISA, including arithmetic, jump, load, and store instructions.  For historical bug information, we  collect 55 historical bug reports from prior work ~\cite{nelson2020specification}. We use the commit title and description of the bug fixes in the linux implementation of eBPF as input and prompt the LLM to group the bugs into high level categories and include a description for each category.  Table~\ref{tab:bug_categories} shows a subset of bug categories generated for eBPF. 

\vspace{1ex}
\noindent\emph{Evaluation Harness.}
We run generated tests using the \texttt{bpf\_conformance} plugin.  
The BPF Conformance project~\cite{bpf-conformance} aims to measure the conformance of  BPF runtimes to the ISA by providing a unified testing interface. The possible test execution outcomes are: 

\begin{itemize}[leftmargin=*,nolistsep]
    \item \textbf{\texttt{PASS}}: ``Test succeeded''. The test is valid and the execution produced the expected return value.  
    \item \textbf{\texttt{FAIL}}: ``Plugin returned incorrect return value x expected y.'' The test is valid but does not pass. This can happen either because there is a bug in the tested implementation, or the LLM generated the incorrect expected output. The differential testing context means that \tool does not rely solely on the LLM to adjudicate expected behavior, instead comparing the output of multiple implementations. 

    \item \textbf{\texttt{SKIP}}: ``Test file contains unsupported instructions/has no BPF instructions.`` The test is not a valid BPF program.
    
    \item \textbf{\texttt{ERROR}}: ``Plugin returned error code 1 and output <msg>.'' The test is again an invalid program producing an error that the conformance plugin can handle, like referencing an invalid register ID in a program instruction.
    
    \item \textbf{\texttt{CRASH}}: ``Unhandled Exception reached the top of main: <msg>.'' The test is invalid in a way that causes the conformance plugin to crash (such as an instruction referencing an invalid label). 
\end{itemize}

\subsection{WebAssembly (Wasm)}
\label{sec:wasm}

\noindent\emph{Background.}
WebAssembly \cite{wasm}, or Wasm, is a portable bytecode that was originally built to run in the browser, but has since become popular in more domains such as cloud and edge computing \cite{wasm-cloud, wasm-edge-cosmonic, wasm-edge-dfinity}, embedded systems \cite{wasm-embedded}, industrial systems \cite{wasm-retrofitting}, and more. 
Wasm provides a safe runtime for untrusted code in many languages (Rust, C/C++, OCaml, and others) as memory is sandboxed; it is also highly performant.  

A Wasm application, consisting of one-to-many \textit{modules}, is run on a Virtual Machine (VM), also referred to as engine or runtime. There are many available implementations including, Wizard Engine~\cite{wizard-gh}, for teaching and research; 
Wasmtime~\cite{wasmtime-gh}, owned by the Bytecode Alliance, for Edge Computing; 
and V8~\cite{v8-gh}, used in Google Chrome, especially noted for its JavaScript integraion.

Before a Wasm module is executed, it is \emph{validated} by the wasm \emph{validator}; this step protects the host system from security vulnerabilities, runtime traps, and undefined behavior.
Generally, each Wasm VM has its own custom Validator, simplifying integration, and improving performance. This motivates standardized testing and robust tooling to ensure runtime conformance.
An established format for testing Wasm runtimes is via \texttt{.wast} tests  written in the human-readable WebAssembly Text (WAT) format.

\vspace{1ex}
\noindent\emph{Tested implementations and code artifacts.} Given the importance of validation to runtime conformance, we focus our testing on the validator modules of four Wasm implementations: (1) Wasm spec \cite{wasm-spec-gh} (the reference implementation, considered the oracle for Wasm behavior), (2) Wizard Engine \cite{wizard-gh}, (3) Wasmtime \cite{wasmtime-gh}, which uses \texttt{wasmparser}'s \cite{wasm-tools-gh} validator under-the-hood, and (4) V8 \cite{v8-gh}. We take tests from the official Wasm test suite~\cite{wasm-spec-gh}.  

Note that \tool was only provided the source code context for the Wasm spec and the Wizard Engine. We test other implementations to deem if the generated tests are useful without further prompting or context.

\vspace{1ex}
\noindent\emph{Natural language artifacts.} We use the Wasm language specification document~\cite{wasm-spec} for testing. Focusing especially  on control-flow instructions, known to be both tricky and error prone, we extract 11 instructions to test from this documentation.  
\tool identifies 11 test files with 596 test cases (from the Wasm spec codebase) for these instructions. We get an initial list of bug categories by prompting ChatGPT (GPT 3.5). 
This list was then verified by a maintainer from Wasm. Table~\ref{tab:bug_categories} shows a subset of bug categories inferred for Wasm.

\vspace{1ex}
\noindent\emph{Evaluation Harness.}
\tool generated tests for Wasm targeting the \texttt{.wast} format (examples in Table~\ref{tab:diff_test_examples_wasm}). For the implementations expecting a different format (like Wizard engine, expecting \texttt{bin.wast}), we use the wasm-spec CLI to translate accordingly and automatically.
Note that invalid or syntactically incorrect tests fail this transformation step. 
Additionally, the tested systems do not all report errors with equal precision or granularity. For consistency, we therefore check for a simple PASS/FAIL result, ignoring the error message produced by test execution. We also refactored multi-assertion test files to provide one assertion per test, for cleaner comparisons of testing results.  We evaluated execution results
by comparing to the Wasm spec reference implementation. 
The possible test execution outcomes are: 
\begin{itemize}[leftmargin=*,nolistsep]
    \item \textbf{\texttt{PASS}}: A validator successfully labeled an invalid module as invalid (as expected by the \texttt{assert}s in the \texttt{.wast}).
    \item \textbf{\texttt{FAIL}}:  The validator labeled an invalid module as valid (as this behavior does not satisfy the \texttt{assert}s in the \texttt{.wast}).
    \item \textbf{\texttt{CRASH}}: An Exception was thrown during execution and printed to the console.
    \item \textbf{\texttt{INVALID}}: The tests fail to convert into desired format. 
\end{itemize}

\vspace{-0.2cm} 

\section{Experimental Design}
\label{sec:setup}

This section describes our experimental setup for evaluating \tool on our systems under test. 
We implement \tool in Python. We make use of GPT-4-32k as the LLM for all the eBPF experiments. Specifically, we use the 0613 version of the GPT-4-32k checkpoint through the Azure OpenAI API. The Wasm experiments have been run using GPT-4o, since the GPT-4-32k model endpoints were deprecated. We use the default values for all the hyper-parameters for both GPT-4-32k and GPT-4o. Specifically, we investigate the following research questions:

\begin{itemize}[leftmargin=*,nolistsep]
\item \textbf{RQ1} How effectively does \tool produce meaningful differential tests? We evaluate this question initially on eBPF (with results for Wasm discussed in RQ3). 
\item \textbf{RQ2} To what extent does each component of \tool contribute to its effectiveness?
\item \textbf{RQ3} How well does \tool generalize across systems?
\end{itemize}

Finally, we qualitatively examine the generated differential tests to determine the cause of differential behavior and possibility of bug attribution.

\subsection{LLM-based Baselines}
\label{sec:llm-baselines}

Both eBPF and Wasm have been extensively tested using traditional fuzzers. However, these do not check for conformance with a specification. We define two baselines that explicitly consider the specification document. First, we use a naive prompting technique that provides the entire specification document and three randomly-selected human tests as context, and prompt the LLM to generate tests for each instruction. Second, as a more targeted baseline, 
we extract the most relevant sections from the document for each instruction. 
We then use the mapping between the instructions and the test files to only use the tests corresponding to the given instruction as examples in the generation prompt. 

We also consider Fuzz4All~\cite{xia2024fuzz4all}, a fuzzing technique that generates and mutates test inputs for projects written in different programming languages. Fuzz4All leverages a larger model, GPT4, to automatically generate prompts for semantically interesting and syntactically valid input, and a cheaper model, StarCoder to generate and mutate these inputs.
To apply Fuzz4All's autoprompting and LLM-fuzzing loop to eBPF, we extend it by adding: (1) an interface to integrate the execution of eBPF bytecode through the Linux libBPF plugin, (2) custom functions to filter, clean, and transform LLM generated output into the required eBPF test format.  
To fuzz eBPF bytecode instructions, we provide Fuzz4All with different sections of the BPF ISA and 3 hand-selected examples of valid bytecode tests for each section. For each section of the ISA, we run experiments with 1000 fuzzing iterations using the best performing configuration reported in the paper. In total, we generate 4,000 tests, using the libBPF plugin to provide evaluation feedback to the auto-prompting and fuzzing loops. Despite having access to few shot examples and instruction semantics from the ISA, 61\% of generated tests did not include valid BPF semantics (e.g., missing exit instructions, hallucinated instruction codes). Considering the remaining 39\% of tests, 38.5\% did not follow the expected test format (e.g. not including --asm or --result), and 0.5\% of tests were successfully parsed but all failed. Upon manual inspection, none of the generated tests passed because the expected result was either empty or invalid. We require valid tests to check for differential behavior, and therefore could not use these generated tests as a baseline. The code and tests generated with Fuzz4All have been included in the artifact.

\subsection{Fuzzing Baselines}
\label{sec:fuzzing-baselines}

We implement two fuzzing baselines to evaluate how LLM-guided test case synthesis (with \tool) compares to more traditional random test case generation. In order to provide a fair comparison, we ensure that both fuzzing baselines are \textit{grammar-aware}, generating \textit{syntactically-valid} eBPF and Wasm programs (but not necessarily semantically-valid).

For the Wasm fuzzer, we use \textsc{Grammarinator}~\cite{hodovan2018grammarinator} to generate syntactically-valid WAT syntax trees according to a reference ANTLR specification. For the eBPF fuzzer, we randomly sample valid eBPF opcodes for instructions and fill their arguments with random values. 
For a fair evaluation, we ran the fuzzer for the same amount of time it took DiffSpec to generate and run tests for both wasm and eBPF. This resulted in 300k tests for wasm run over $\sim6$ hours and 200k tests for eBPF run over $\sim12$ hours.

\subsection{Evaluation Metrics (RQ1)}

We use the following metrics in evaluating \tool performance:

\noindent\textbf{Validity.} A generated test is \textit{valid} if it results in \texttt{PASS}, \texttt{FAIL}, or \texttt{ERROR} across all implementations --- that is, \tool produced a syntactically valid test. Tests that are skipped or lead to crashes are considered invalid. 

\noindent\textbf{Differentiating tests.} A \textit{differentiating test} is one that produces a different outcome/return value for at least two different implementations. For example, a test resulting in \texttt{PASS} on one system and \texttt{FAIL} or \texttt{ERROR} on another, or a test resulting in \texttt{FAIL} on two systems but return different values that do not match the test's expected value. 
Note that it is still possible for the generated expected value to be incorrect. However, this risk is mitigated by the use of multiple systems in the differential testing context, where their outputs can be compared independently of the test's oracle value.

\noindent\textbf{Test complexity.}  Automatically generated tests can be difficult for humans to understand or use~\cite{gen-and-maintenance}. This can be especially risky in the differential testing context~\cite{csmith, naturalness}.  
We use the number of lines in the tests as a simple proxy for test complexity.

\noindent\textbf{Test diversity.} To assess the ability of \tool to generate a \textit{diverse} set of tests covering many different behaviors, we examine the generated tests and compute the unique number of features across the set. For eBPF we consider: unique instructions (e.g. eBPF opcodes), unique registers (e.g. \texttt{\%r0}, \texttt{\%r1}, ...), unique memory addresses (e.g. \texttt{[\%r3]} or \texttt{[\%r1+4]}), and unique immediate values (e.g. \texttt{0x1}, \texttt{0x1337}, ...).

\subsection{Configurations for ablation (RQ2)}

We evaluated the performance of \tool under multiple configurations (summarized in Table~\ref{tab:results-ebpf}) to evaluate the contribution of the context extracted from the different artifacts, and the effect of the two-stage test generation procedure. 
These settings are:

\noindent\textbf{3-shot-random}: Our first baseline; uses the instruction (\instruction), the language specification (\fullSpec), and three random test examples (\tests) to directly generate tests (\testCode). 

\noindent\textbf{target-section}:  Our second baseline; uses the instruction (\instruction), the manual section under which the instruction was listed (\relevantSpec), and targeted examples (\tests) from the test suite to directly generate tests(\testCode).


\noindent\textbf{prompt-chain}: Asks an LLM to extract the key information about an instruction from the manual (\constraints), and uses a two-step test generation process, combined using prompt chaining.  We include three random test examples (\tests)  when generating the test (\testCode) from the test descriptions (\testDesc). 

\noindent\textbf{prompt-chain-instruct}: Provides a curated set of guidelines to ensure valid tests, such as ``ensure that the output is in \%r0'' for eBPF. We provide these guidelines (\guidelines) during test generation, on top of the \textbf{prompt-chain} approach. 

\noindent\textbf{bug-centric}: Builds on prompt-chain-instruct setup by additionally providing the LLM with cues from historical bug data (\bugClasses). 

\noindent\textbf{code-description}: Builds on prompt-chain-instruct setup by additionally extracting and summarizing relevant code snippets (\codeDesc). 

\noindent\textbf{code-description-diff}: Builds on code-description by generating a list of implementation differences (\codeDescDiff) in the two code descriptions. 

\noindent\textbf{code-diff}: We observed that using the differences in descriptions (\codeDescDiff) can help the LLM generate differentiating tests. In this setup, we have the LLM generate a list of differences directly from the relevant code snippets (\codeDiff) extracted from the implementations. We then use each difference that was generated to guide the LLM when generating test descriptions (uses two step test generation).

\noindent\textbf{bug-guided-code-diff}, or \tool: Combines all useful elements: prompt-chain-instruct (\constraints  ~ \guidelines ~ \tests) with bug categories (\bugClasses) and code differences (\codeDiff), with a two step generation process (\testDesc ~\testCode).

Note that we conducted these experiments first on the eBPF system; we then used the configuration that led to the most differential tests to subsequently test Wasm validators.

\begin{table*}[ht!]
\centering
    \caption{Performance of \tool compared against baselines (first 3 rows) and various ablations on eBPF runtimes. Legend: instruction: \instruction, specification document: \fullSpec, relevant section of specification document: \relevantSpec, constraints: \constraints,  example tests: \tests, test descriptions: \testDesc, guidelines: \guidelines, bug classes: \bugClasses, code descriptions: \codeDesc, code descriptions diff: \codeDescDiff, code diffs: \codeDiff, W: windows, L: linux, U: ubpf, \ensuremath{\dagger}: fuzzer generates all possible instruction variants and millions of unique addresses/values } 
    \label{tab:results-ebpf}

\begin{tabular}{l|l|r|rrrr|rrrr}

\toprule
\multirow{2}{*}{\textbf{Ablation type}} & \multirow{2}{*}{\textbf{Context used}} & \multirow{2}{*}{\textbf{Validity (\%)}} & \multicolumn{4}{c|}{\textbf{\textit{Test Diversity (Unique \#)}}} & \multicolumn{4}{c}{\textbf{\textit{Differential Tests Found}}} \\


 & &  \textbf{} & \textbf{Instr.} & \textbf{Reg.} & \textbf{Addr.} & \textbf{Values} & {\textbf{W-L}} & {\textbf{L-U}} & {\textbf{W-U}} & {\textbf{Total}}  \\
 
 \midrule
fuzzing & \textcolor{black}{\instruction} & 100.0 & $\dagger$ & $\dagger$ & $\dagger$ & $\dagger$ & 1 & 1 & 2 & \textbf{4} \\

3shot-random & \textcolor{black}{\instruction} ~ \textcolor{black}{\fullSpec} ~ \textcolor{black}{\tests} ~  \textcolor{white}{\testDesc} ~ \textcolor{white}{\guidelines} ~ \textcolor{white}{\bugClasses} ~ \textcolor{white}{\codeDiff}  & 68.3 & 25 & 2 & 0 & 79 & 14 & 13  &  4 & \textbf{14} \\

target-section & \textcolor{black}{\instruction} ~ \textcolor{black}{\relevantSpec} ~ \textcolor{black}{\tests} ~  \textcolor{white}{\testDesc} ~ \textcolor{white}{\guidelines} ~ \textcolor{white}{\bugClasses} ~ \textcolor{white}{\codeDiff}  & 76.3 & 52 & 8 & 14 & 91 & 39 & 37  & 3  & \textbf{39}  \\\midrule

prompt-chain & \textcolor{black}{\instruction} ~  \textcolor{black}{\constraints} ~ \textcolor{black}{\tests} ~ \textcolor{black}{\testDesc} ~ \textcolor{white}{\guidelines} ~ \textcolor{white}{\bugClasses} ~ \textcolor{white}{\codeDiff}  & 24.8 & 34 & 12 & 2 & 118 & 23 & 21  & 6  & \textbf{24}  \\

prompt-chain-instruct & \textcolor{black}{\instruction} ~  \textcolor{black}{\constraints} ~ \textcolor{black}{\tests} ~ \textcolor{black}{\testDesc} ~ \textcolor{black}{\guidelines} ~ \textcolor{white}{\bugClasses} ~ \textcolor{white}{\codeDiff} & 66.3 & 46 & 13 & 5 & 181 & 38 & 34 & 9 & \textbf{38} \\

bug-centric & \textcolor{black}{\instruction} ~ \textcolor{black}{\constraints} ~ \textcolor{black}{\tests} ~  \textcolor{black}{\testDesc} ~ \textcolor{black}{\guidelines} ~ \textcolor{black}{\bugClasses} ~ \textcolor{white}{\codeDiff} & 64.8 & 64 & 15 & 26 & 500 & 266 & 251 & 53 & \textbf{271}   \\

code-description & \textcolor{black}{\instruction} ~  \textcolor{black}{\constraints} ~ \textcolor{black}{\tests}  ~ \textcolor{black}{\testDesc} ~ \textcolor{black}{\guidelines} ~ \textcolor{white}{\bugClasses} ~ \textcolor{black}{\codeDesc} & 63.4 & 45 & 13 & 8 & 189 & 29 & 29  & 3 & \textbf{29} \\

code-description-diff & \textcolor{black}{\instruction} ~ \textcolor{black}{\constraints} ~ \textcolor{black}{\tests} ~ \textcolor{black}{\testDesc} ~ \textcolor{black}{\guidelines} ~ \textcolor{white}{\bugClasses} ~ \textcolor{black}{\codeDescDiff} & 65.3 & 63 & 13 & 25 & 446 & 158 & 137 &  32 & \textbf{159} \\

code-diff & \textcolor{black}{\instruction} ~ \textcolor{black}{\constraints} ~ \textcolor{black}{\tests} ~ \textcolor{black}{\testDesc} ~ \textcolor{black}{\guidelines} ~ \textcolor{white}{\bugClasses} ~ \textcolor{black}{\codeDiff} & 66.5 & 61 & 16 & 23 & 404 & 198 & 139 & 79 & \textbf{200} \\\midrule

\tool: bug-guided-code-diff & \textcolor{black}{\instruction} ~ \textcolor{black}{\constraints} ~ \textcolor{black}{\tests} ~ \textcolor{black}{\testDesc} ~ \textcolor{black}{\guidelines} ~ \textcolor{black}{\bugClasses} ~ \textcolor{black}{\codeDiff} & 69.4 & 87 & 16 & 106 & 1850 & 1886 &  1790 & 226 & \textbf{1901} \\\bottomrule
 
\end{tabular}
\end{table*}

\section{Results}
\label{sec:results}

This section presents experimental results, speaking to \tool effectiveness (Section~\ref{sec:rq1}); the contributions of individual design decisions and types of information to that effectiveness (Section~\ref{sec:rq2}); and \tool's generalizability to diverse domains (Section~\ref{sec:rq3}).

\subsection{RQ1: Overall effectiveness}
\label{sec:rq1}

Table~\ref{tab:results-ebpf} summarizes the results of running \tool on the eBPF systems, as discussed in Section~\ref{sec:ebpf}.  The first three rows of the tables show baselines; the final row, the results for \tool (the intermediate rows are discussed in Section~\ref{sec:rq2}).

The fuzzing baseline is designed to generate \textit{syntactically valid} tests and therefore has a 100\% validity. The two LLM-based baselines also generate tests with high validity (68\% and 76\%) since they primarily test for simple cases. \tool's integration of bug category and code differences improves test validity (69\%) over other ablations,
while identifying many more behavioral differences. 

Additionally, looking at number of differential tests found, we see that despite both fuzzing and LLM-based baseline approaches having high validity rates, they generate very few differential tests (4, 27, and 75). 
In contrast, \tool identifies 1901 differential tests, while having comparable validity. We manually analyze a random sample of 200 differential tests to identify potential bugs in eBPF. Note that there can be multiple differentiating tests pointing to the same underlying bug. Using this analysis, we identify 4 concrete bugs for eBPF, and have filed reports with the maintainers of the associated implementations. All of them have been confirmed as real bugs.
For eBPF, we observe the following classes of tests that demonstrate meaningful behavioral differences among different implementations; Table~\ref{tab:diff_test_examples_ebpf}. shows examples of tests corresponding to these categories, which are: 

\begin{itemize}[leftmargin=*,nolistsep]

\item Uses uninitialized registers, includes not storing  output in r0.

\item Undefined behaviour when using stack pointers (\%r10).
    
\item Tests resulting in potential memory leaks.
\item Inconsistencies in how jump instructions are handled.

\item Tests containing call instructions to helper functions (differential behavior is expected).

\item Tests with infinite loops that causes the ebpf-for-windows verifier to hang. 

\end{itemize}

\begin{table}[ht]
\small
  \begin{center}
    \caption{Example differentiating tests generated by \tool for eBPF runtimes that identified bugs}
    \label{tab:diff_test_examples_ebpf}
    
    \begin{tabular}{p{3cm}p{4cm}p{5cm}}

    \toprule
    \textbf{Generated Test} & \textbf{Execution Outcomes}\\
    \midrule
     \texttt{-- asm\newline 
    ldxw \%r0, [\%r1]\newline 
    exit\newline
    -- mem \newline 
    00 00 00 00\newline
    -- result\newline 
    0x0} & \textit{(Kernel Memory Leak)}\newline
    \underline{Windows:} PASS: Test succeeded\newline \underline{uBPF:} PASS: Test succeeded\newline \underline{Linux:} FAIL: Plugin returned incorrect return value ffff8b09dc604100 expected 0\\
    \midrule
    
    \texttt{-- asm\newline
    mov \%r1, 5\newline
    jset \%r1, \%r1, lbl1\newline
    mov \%r0, 0\newline
    exit\newline
    lbl1: mov \%r0, 1\newline
    exit\newline
    -- result\newline
    0x1} & \textit{(Inconsistencies in jump)}\newline
    \underline{Windows:} FAIL: Plugin returned incorrect return value 0 expected 1\newline \underline{uBPF:} FAIL: Plugin returned incorrect return value 7ffff338a820 expected 1\newline \underline{Linux:} ERROR: Plugin returned error code 1\\

   \bottomrule      
    \end{tabular}
  \end{center}
\end{table}

\begin{highlight}
Using \tool, we were able to generate differential tests that uncovered four different bugs in the different implementations of eBPF, despite many of these being extensively tested by fuzzers and other techniques.  These bugs have been confirmed by the maintainers of the bpf conformance project and are currently in the process of being fixed.
\end{highlight}

\begin{figure}[ht]
\centering
\includegraphics[width=0.45\textwidth]{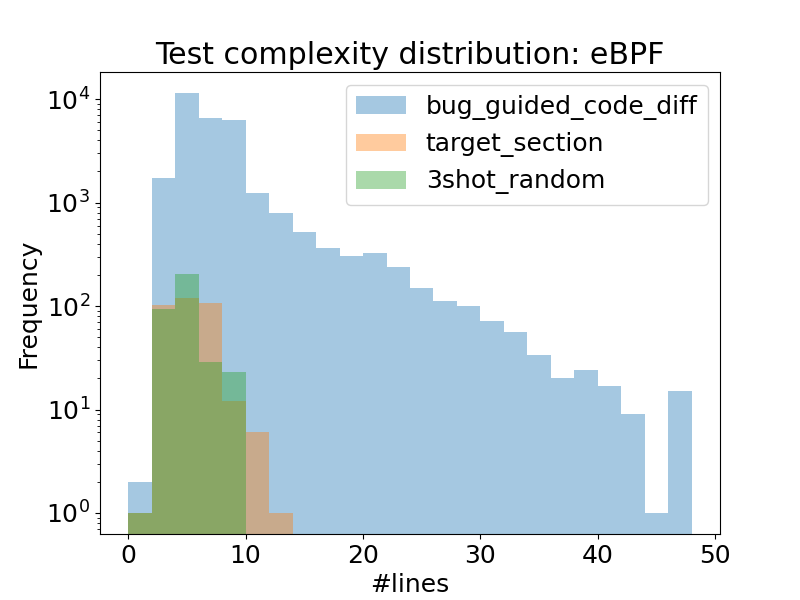}
  \caption{The distribution of generated test complexity (measured by test length) for eBPF. The bug-guided-code-diff (\tool) generates a more complex distribution of tests.} 
  \label{fig:test-complexity}
\end{figure}

\vspace{1ex}
\noindent\emph{Test complexity.} Figure~\ref{fig:test-complexity} shows the distribution of test complexity of the tests generated by \tool for eBPF. We observe similar trends for Wasm. For eBPF, we additionally compare the complexity of the tests generated by \tool with the two baseline approaches. We find that overall, \tool generates short tests that average fewer than 20 lines. Additionally, we observe that the tests generated by the baseline approaches are much shorter than the ones that \tool generates, which highlights that \tool generates more complex tests that can find differentiating behavior in different implementations. Despite the baselines having a higher validity rate, these valid tests are not effective at finding interesting differential behavior.

\vspace{1ex}
\noindent\emph{Test diversity.} Our test diversity results (Table~\ref{tab:results-ebpf}) show an interesting story. Both the LLM baselines have lower unique numbers of program features such as types of instructions and registers. As we include more context, the resulting test cases cover more of the possible eBPF instruction variants and test more modes of operation (such as using different memory addresses), correlated with an increase in the number of differentiating test cases. Interestingly however, the fuzzing baseline covers far more possible program variants (by orders of magnitude), yet finds the \textit{fewest} number of differentiating tests. These results suggest that syntax-adherence and diversity alone is insufficient to find interesting tests, and \textit{guidance}, for example though the use of specification, source code, guidelines, etc., is \textit{critical} to exercise interesting behavior.

\begin{figure*}
    \center
    \includegraphics[width=\textwidth]{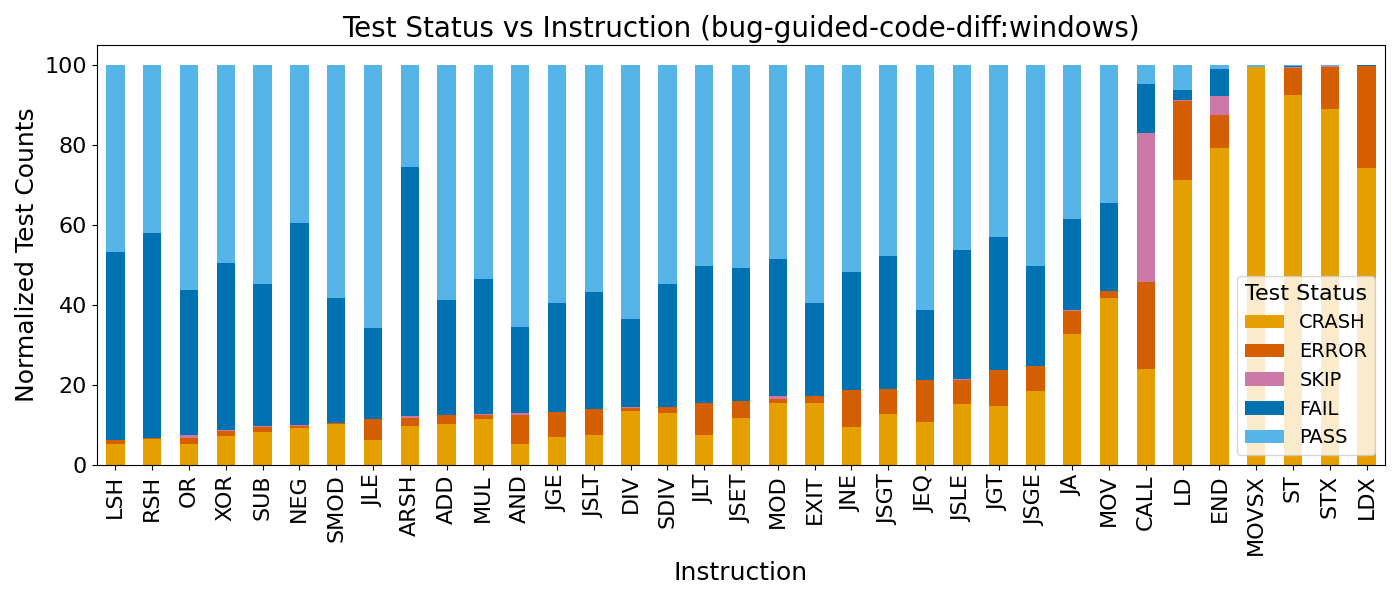}
    \vspace{-4mm}
    \caption{Visualization of test status distribution per instruction for eBPF using \tool on Windows.}
 
    \label{fig:instruction-test-status}
  \end{figure*}

\vspace{1ex}
\noindent\emph{Instruction-level performance}
Figure~\ref{fig:instruction-test-status} shows the distribution of the test status or execution outcome of all the generated tests for each instruction using the bug-guided-code-diff approach on the windows implementation of eBPF. Upon closer analysis, we find that the \tool generates mostly valid tests (that PASS or FAIL) for most arithmetic and logic instructions. The percentage of valid tests declines for more complex instructions like the jump instructions. On the other hand, load/store instructions, namely, \texttt{LD}, \texttt{ST}, \texttt{LDX}, \texttt{STX}, along with \texttt{END} and \texttt{MOVSX}, have the highest invalid test rate (tests either CRASH or throw an ERROR). This gives us insights into the types of instructions that LLMs can and cannot reason about. We observe similar trends when the generated tests are run on other implementations of eBPF using the different ablations. Additional plots for different ablations and different implementations can be found in the supplementary material.

\paragraph{Note on specification and code coverage} We manually confirmed the model successfully extracted (and generated tests for) all instructions in each language. We also manually confirmed all specifications related to a given instruction were correctly extracted for a small sample. These judgments informally suggest good specification coverage. Doing this completely is likely infeasible since the specifications are long. Given the differential testing goal (to find bugs by comparing implementations), we consider observed differential behavior a more instructive metric than code coverage. We don’t expect much coverage improvement, for such well-tested projects, with a technique that doesn’t target it.

\subsection{RQ2: Ablations}
\label{sec:rq2}
\label{sec:ablation}

Table~\ref{tab:results-ebpf} also summarizes the results of the  ablations for eBPF. 
As above, targeted prompting has the highest validity rate, but the generated tests are often simple. 
Meanwhile, comparing prompt-chain with prompt-chain-instruct, we find that the human written instructions/ guidelines helps improve validity by a huge margin of ~42\%. 
Bug context alone generates complex tests, and having code context (in all forms) helps improve the validity of the tests. On the other hand, using the bug categories and differences in code are extremely useful for generating differentiating tests, with the number of differentiating tests going up to 271 and 200 respectively. 

Overall,  most ablations that only look at the natural language specifications document and existing tests generate fewer than 40 differentiating tests. Interestingly, while the differences in code descriptions are seemingly useful, with 159 differentiating tests generated, the code descriptions on their own are not very useful, resulting in only 29 differentiating tests. 

\begin{highlight}
The bug and code diff context helps generate more complex tests that identify differential behavior in different implementations. Guiding the model with bug categories, and differences in code are especially useful. A two-step test generation process with the human written instructions is more effective. 
\end{highlight}

\subsection{RQ3: Generalizability to Wasm}
\label{sec:rq3}

We  demonstrate \tool's ability to generalize beyond eBPF by evaluating its performance on Wasm validators.  
We use the best combination of features, substantiated by the ablation study, for these experiments.  
With respect to validity, using the Wasm spec's translation facilities as a proxy, 85\% of the generated tests were valid (could be translated). 

\begin{table}[ht]
\centering
    \caption{\tool vs. fuzzing baseline performance on Wasm validators, compared to the Wasm spec reference implementation.}
    \label{tab:results-wasm}
\begin{tabular}{l|rr}
\toprule
\multirow{2}{*}{\textbf{Comparison}} & \multicolumn{2}{c}{\textbf{\textit{Differential Tests Found}}} \\
& \textbf{\tool} & \textbf{Fuzzing baseline}\\ \midrule
Wizard Engine vs. Wasm spec & 6 & 74 \\ 
Wasmtime vs. Wasm spec & 256 & 0 \\ 
V8 vs. Wasm spec & 37 & 0 \\\midrule
Total & 299 & 74\\ 
\bottomrule
\end{tabular}
\end{table}
Table \ref{tab:results-wasm} summarizes the remaining results. We find a total of 74 differentiating tests with the fuzzing baseline. However, upon manual inspection, we found that the Wizard engine imposes a hard limit on memory and table size (capped at 10000000) during validation, whereas the spec interpreter does not. Therefore, differential behavior is expected, finding no underlying bug.
On the other hand, \tool produced a total of 299 differentiating tests across the four different implementations of Wasm validators.
Upon manual analysis, we observe the following classes of tests that demonstrate behavioral differences among different implementations for Wasm validators, and therefore highlight potential bugs:

\begin{itemize}[leftmargin=*,nolistsep]
    \item Tests with type mismatch (expected vs. actual type).
    \item Tests with invalid type (extremely large numbers).
    \item Implementations had different semantics for \texttt{.wast} assertions. The \texttt{assert\_malformed} and \texttt{assert\_invalid} cases are treated differently on the Wasm spec, but the same on all other implementations (differential behavior is expected).
    \item Implementations had different semantics for handling unavailable imports, error vs. validate what is available (differential behavior is expected).
\end{itemize}

The two bugs we identified, (i) \emph{Type Mismatch}: Validate that \texttt{[return\_]call\_indirect} operates on a table with \texttt{funcref},
and (ii) \emph{Unknown Type}: Fix cast of out-of-bounds values, 
have been reported to project maintainers, and have since been fixed.  Table~\ref{tab:diff_test_examples_wasm} shows the tests, which have also been added to test suite of both Wizard and wasm-spec.

\begin{table}[ht!]
\small
  \begin{center}
    \caption{Differentiating tests generated by \tool for Wasm validators that identified bugs.}
    
    \label{tab:diff_test_examples_wasm}
    
    \begin{tabular}{p{5.5cm}p{2.6cm}}

    \toprule
    \textbf{Generated Test} & \textbf{Execution Outcomes}\\
     
     \midrule
    \texttt{(assert\_invalid (module (type (func))\newline
     (table 10 externref)\newline
     (func \$call-indirect (call\_indirect\newline
     (type 0) (i32.const 0)))) "type mismatch")}
     & \textit{(Type mismatch)}\newline
     wasm-spec: PASS\newline wizard-engine: FAIL \\
     
    \midrule
    \texttt{(assert\_invalid 
  (module(type (func (param i32)))\newline
    (table 1 funcref)\newline
    (func \$conditional-dangling-type\newline
      (if (i32.const 1)\newline
        (then (call\_indirect (type 0xffffffff) (i32.const 0))))))
  "unknown type")}
     & \textit{(Unknown type)} \newline
     {wasm-spec:} PASS\newline {wizard-engine:} CRASH \\
    
   \bottomrule      
    \end{tabular}
  \end{center}
\end{table}

\begin{highlight}
The differential tests generated by \tool uncovered 2 different bugs in the Wizard Engine validator, despite being extensively tested. These bugs were reported to the maintainers and have now been fixed.  This speaks to \tool's ability to differentially test a variety of systems.
\end{highlight}

\section{Related Work}

\noindent\emph{Testing eBPF.}
There exists a large body of work on improving eBPF verifiers and JIT compilers through fuzzing~\cite{hung2024brf, vyukov2015syzkaller, iovisor2015, li2023fuzzing}, state embeddings~\cite{sun2024validating}, or rewriting the verifier with proof-carrying code~\cite{nelson2020specification, vishwanathan2023verifying, jin2024enhanced}, including using abstract interpretation to prove various functional and safety properties~\cite{gershuni2019simple}. However, as existing eBPF ecosystems continue to grow in complexity and novel runtimes are added~\cite{nelson2020specification, bpfwindows}, concerns regarding correctness remain. Despite extensive testing and verification efforts, kernel bugs introduced by the verifier and JIT, as well as exploits leveraging unsafe extensions that pass the verifier but violate other safety properties, are constantly reported~\cite{jia2023kernel}.

Kernel fuzzing techniques like LKL-fuzzer~\cite{mohamed2023understanding}, BRF~\cite{hung2024brf}, or BVF~\cite{sun2024finding}, generate eBPF programs that passthe verifier to find correctness bugs, using structured program generation to enforce the eBPF ISA grammar. 
\tool instead relies on LLMs to infer the grammar and generate valid bytecode from the specification document and example tests, overcoming common limitations of generation-based fuzzing techniques, namely that they do not evolve with the program semantics, and they have restricted generation ability. 
Closer to our work is Kgent~\cite{zheng2024kgent}, which uses LLM agents to generate valid eBPF programs grounded in formal specifications generated from natural language documentation. While we also seek to leverage informal natural language specifications, we focus on generating executable tests with the purpose of identifying differential behavior.

\vspace{1ex}
\noindent\emph{Testing Wasm.}
There has been work to further the correctness guaranteed by the Wasm spec through mechanization via a custom DSL~\cite{spec-tech}.
Naturally, bugs still exist due to gaps in testing, diversity in use cases, variation in implementation, etc., and have been studied~\cite{wasm-bug-empirical, wasm-bug-compilers}. 
Tools to find such bugs use static and dynamic analysis~\cite{wasabi,tough-call,wizard-inst,wasm-aop,wasma,eunomia}, compiler fuzzing~\cite{wasm-smith, wa-fuzzer, wasmcfuzz} and binary fuzzing~\cite{wafl, fuzzm, wapplique}. Some fuzzers have been developed to target the behavior in a specific domain such as smart contracts~\cite{wasai}.

Efforts to use differential testing for Wasm have taken several forms (such as using a stack-directed binary generator~\cite{stack-driven}), none of which (as far as we are aware) use LLMs or leverage natural language. 
DITWO~\cite{ditwo} leveraged differential testing to uncover missed Wasm optimization opportunities. WADIFF~\cite{wadiff}, the first differential testing framework for Wasm, generated test cases for each operator and then fuzzed them. 
Wasmaker~\cite{wasmaker} performs similarly, but can generate more complex binaries.

\noindent\emph{Testing with LLMs}
Neural test generation techniques have been developed to address limitations and challenges of traditional testing techniques, for example, improving coverage of the input space and readability of the generated tests. Several works rely on LLMs to improve unit test generation techniques, by leveraging build ~\cite{alshahwan2024automated} and code coverage information~\cite{pizzorno2024coverup, lemieux2023codamosa}, using multi-step prompting with AST-based context retrieval \cite{ryan2024code}, and training models on aligned code and tests to improve generated test validity~\cite{rao2023cat}.

Fuzzing techniques, like TitanFuzz~\cite{deng2023large} which uses LLMs to generate and mutate human-like code to test deep learning library APIs, have also been used to improve the coverage and quality of fuzzing inputs.
 Fuzz4All~\cite{xia2024fuzz4all} aims to address the limitations of traditional compiler fuzzers by leveraging LLMs as an input generation and mutation engine. Fuzz4All relies on a set of user provided documentation, example code, or formal specifications for each component under test. It then uses autoprompting techniques to summarize these artifacts and iteratively mutate generated inputs. Our approach leverages similar inputs, however \tool does not require users to manually extract relevant sections of documentation and other artifacts for the component under test. 
 Instead, \tool automatically extracts relevant specifications for each instruction from the given document. This is to ensure higher level specifications, which can be at scattered across a document, are not missed, improving the validity of the generated inputs.   While TitanFuzz and Fuzz4All generate input programs for a single system, \tool generates test inputs that differentiate two given programs, does not require a user defined oracle, and uses prompt chaining to incorporate evolving differential information, such as difference in code implementations and historic bugs, to guide test generation. 
The target-section baseline we use closely resembles the Fuzz4All approach. 
Closest to our work is Mokav~\cite{etemadi2024mokav}, an LLM-guided  differential testing technique. Mokav uses execution based feedback to prompt models to generate difference exposing tests between two versions of a python program. Similar to our work, Mokav generates natural language descriptions of each versions of the program, however, unlike our approach it does so independently, without prompting the model to explicitly look for differences. 
While Mokav targets differential testing, it does not consider other sources of natural language artifacts to guide test generation. Other approaches, such as AID~\cite{liu2024llm} and a Differential Prompting framework introduced by Li et al.~\cite{li2023nuances}, leverage buggy versions of code to generate fault-localizing tests that expose differences between a buggy and fixed version. In contrast, \tool directly generates differentiating tests without access to a known buggy version of the code.

\section{Conclusion}
\label{sec:discussion}

In this paper, we present \tool, a novel approach to differential testing that is driven by both natural language specifications and code artifacts. \tool harnesses large language models to learn from vast amounts of natural language specifications, source code, and historical bug data, enabling it to generate targeted tests that reveal meaningful differences between the systems under test. We evaluate our approach on multiple implementations of two extensively tested and widely adopted frameworks: Wasm and eBPF runtimes. \tool successfully generates over $\sim2200$ differential tests across both systems that uncover four distinct, previously unknown, bugs in eBPF runtimes and at least two confirmed, now-fixed, bugs in Wasm validators. Our findings demonstrate that leveraging large language models to generate tests based on semantic differences between program versions can be a highly effective technique for differential testing.
Future work involves incorporating execution feedback to our prompt-chaining approach to refine tests, and applying \tool to other systems like language compilers, EVM, or network protocol parsers.

\noindent\textbf{Data Availability}: All the code for \tool along with the prompts to generate the tests, the scripts to run the tests on the evaluation harness, the generated differential tests, results from additional experiments and manual analysis are available at: \url{https://doi.org/10.5281/zenodo.13756137}.


\balance

\bibliographystyle{ACM-Reference-Format}
\bibliography{references}

\balance

\newpage\clearpage

\end{document}